\documentclass[prl,preprint,showpacs]{revtex4} 
\input{epsf}  
\begin{document}  
\title{Efficient simulations of gas-grain chemistry
in interstellar clouds}  
\author{Azi Lipshtat and Ofer Biham}  
\affiliation{  
Racah Institute of Physics,   
The Hebrew University,   
Jerusalem 91904,   
Israel}  
 
\begin{abstract}  

Chemical reactions on dust grains 
are of crucial importance in 
interstellar chemistry
because they  
produce molecular hydrogen and
various organic molecules.
Due to the submicron size of the grains and the low flux,
the surface populations of reactive species 
are small and strongly fluctuate.
Under these conditions rate equations fail
and the master equation is needed 
for modeling these reactions.
However, the number of equations in the master equation grows
exponentially with the number of reactive species, 
severely limiting its feasibility.
Here we present a method which dramatically reduces the number of
equations, 
thus enabling the incorporation of the master equation 
in models of interstellar chemistry.

\end{abstract}  
  
 
\maketitle  

The chemistry of interstellar clouds consists of reactions taking place in 
the gas phase as well as on the surfaces of dust grains 
\cite{Hartquist1995}. 
Surface reactions include the formation 
of molecular hydrogen 
\cite{Gould1963,Hollenbach1970}
%,Hollenbach1971a,Hollenbach1971b}, 
and reaction networks producing  
various organic molecules. 
%Hydrogen molecules are a necessary component 
%for the initiation of gas-phase reaction 
%networks that give rise to the chemical complexity observed in interstellar   
%clouds. 
The simulations of 
grain-surface chemistry
%chemical reaction networks on interstellar dust grains 
are typically done using rate 
equation models 
\cite{Pickles1977,Hendecourt1985,Brown1990b,Hasegawa1992,Caselli1993}.  
These models consist of coupled ordinary differential equations that 
provide the time derivatives of the populations of the 
reactive species on the grains.
Rate equations are highly efficient for the simulation of
reactions on macroscopic surfaces.  
However, in the limit of small grains under low flux,  
rate equations fail because they ignore the 
discrete nature of the populations of reactive species
and their fluctuations
\cite{Tielens1982,Charnley1997,Caselli1998,Shalabiea1998,Stantcheva2001}.  
%For example, as the number of H atoms on a 
%grain fluctuates in the range of 0, 1 or 2, 
%the formation rate 
%of H$_2$ molecules
%cannot be obtained from the average  
%number alone. 
%This can be easily understood, since the 
%recombination process requires at least two 
%H atoms simultaneously on the surface. 
 
Recently, a master equation approach 
for the simulation of reaction networks on small grains  
was proposed
\cite{Biham2001,Green2001}. 
It takes into account the discreteness and fluctuations
in the surface populations of reactive species
and provides accurate results for the reaction rates.
For example,
in the case of hydrogen recombination,  
its dynamical variables are the 
probabilities $P(N)$  
that there are  
$N$  
hydrogen
atoms 
on a grain. 
The time derivatives   
$\dot P(N)$,  
$N = 0, 1, 2, \dots$, 
are expressed  
in terms of the adsorption, reaction and desorption terms. 
The master equation 
was applied to the study of
reaction networks involving multiple species 
that appear in models of interstellar chemistry
\cite{Stantcheva2002,Stantcheva2002b}.
The master equation can be solved either by direct integration
\cite{Biham2001,Green2001} 
or by using a Monte Carlo (MC) method
\cite{Charnley2001}. 
A significant advantage of 
direct integration over the MC approach
is that the equations
can be easily coupled to the  
rate equations of gas-phase chemistery. 
However, the number of coupled equations increases exponentially 
with the number of reactive species, 
making direct integration infeasible for complex reaction
networks of multiple species
\cite{Stantcheva2002,Stantcheva2002b}.

In this paper we introduce  
the multi-plane method,
in which the number of equations 
is dramatically reduced, 
thus enabling the incorporation of the master equation
in models of interstellar chemistry.
The multi-plane method is tested by comparing its results  
to those obtained from the complete master equation set,  
showing excellent agreement. 
To demonstrate the method we consider a reaction network that
involves three reactive species: H and O atoms and OH molecules
\cite{Caselli1998,Shalabiea1998,Stantcheva2001}.
For simplicity we denote the reactive species by
$X_1=$ H,
$X_2=$ O,
$X_3=$ OH,
and the resulting non-reactive species 
by
$X_4=$ H$_2$,
$X_5=$ O$_2$,
$X_6=$ H$_2$O.
The reactions that take place in this network include
H + O 
$\rightarrow$ 
OH
($X_1 + X_2 \rightarrow X_3$),
H + H 
$\rightarrow$ 
H$_2$
($X_1 + X_1 \rightarrow X_4$),
O + O 
$\rightarrow$ 
O$_2$
($X_2 + X_2 \rightarrow X_5$),
and
H + OH 
$\rightarrow$ 
H$_2$O
($X_1 + X_3 \rightarrow X_6$).
  
Consider a small spherical grain of diameter $d$,  
exposed to fluxes  
of H and O atoms and OH molecules.
The cross-section of the grain is 
$\sigma=\pi d^2/4$ 
and its surface area is $\pi d^2$. 
The density of adsorption sites on the surface 
is denoted by 
$s$ (sites cm$^{-2}$). 
Thus, the number of  
adsorption sites on the grain is 
$S=\pi d^2 s$. 
The desorption rates of atomic and molecular species on the
grain are given by   
$W_i =  \nu \cdot \exp [- E_{1}(i) / k_{B} T]$,    
where $\nu$ is the attempt rate   
(standardly taken to be $10^{12}$ s$^{-1}$),   
$E_{1}(i)$   
is the activation energy barrier for desorption   
of specie
$X_i$
and $T$ (K)
is the surface temperature.  
The hopping rate of adsorbed atoms between 
adjacent sites on the surface is 
$a_i =  \nu \cdot \exp [- E_{0}(i) / k_{B} T]$,   
where
$E_{0}(i)$ is the activation energy barrier for hopping 
of $X_i$ atoms (or molecules).  
Here we assume that diffusion occurs only by thermal hopping,  
in agreement with experimental results  
\cite{Katz1999}.  
For small grains, it is convenient to replace the  
hopping rate $a_i$ (hops s$^{-1}$) by 
the sweeping rate
$A_i = a_i / S$, 
which is approximately the inverse of the time it takes  
for an $X_i$ atom to visit nearly all  
the adsorption sites on the grain surface
(up to a logarithmic correction 
\cite{Krug2003}). 

In the reaction network described above, the dynamical variables
of the master equation are the probabilities
$P(N_1,N_2,N_3)$
of having a population that includes $N_i$ atoms 
of specie $X_i$ on the grain. 
Only the reactive species are included in the master equation,
from which the production rates of the
non-reactive species can be obtained.
The master equation 
for the H, O and OH system takes the form
\begin{eqnarray}  
&&\dot P(N_1,N_2,N_3) = 
\sum_{i=1}^3  
F_i \left[ P(..,N_i-1,..) - P(N_1,N_2,N_3) \right]  
\nonumber \\ 
&&+\sum_{i=1}^3  
W_i \left[ (N_i+1) P(..,N_i+1,..) - N_i P(N_1,N_2,N_3) \right]  
\nonumber \\ 
&&+ \sum_{i=1}^2  
A_i \left[ (N_i+2)(N_i+1) P(..,N_i+2,..) 
           - N_i(N_i-1) P(N_1,N_2,N_3) \right]  
\nonumber\\ 
&&+ (A_1+A_2) \left[ (N_1+1)(N_2+1) P(N_1+1,N_2+1,N_3-1) 
           - N_1 N_2 P(N_1,N_2,N_3) \right] 
\nonumber\\ 
&&+ (A_1+A_3) \left[ (N_1+1)(N_3+1) P(N_1+1,N_2,N_3+1) 
           - N_1 N_3 P(N_1,N_2,N_3) \right]. 
\label{eq:Master}  
\end{eqnarray}  
 
\noindent  
The terms in the first sum describe the incoming flux,  
where $F_i$ (atoms s$^{-1}$)   
is the flux {\em per grain}  
of the specie $X_i$.  
The probability 
$P(..,N_i,..)$ 
increases   
due to adsorption of an $X_i$ atom on grains that already  
have $N_i-1$ such atoms on their surfaces,
and decreases 
due to adsorption on grains that have
$N_i$ atoms.
Similarly, the second sum describes the effect of desorption.   
The third sum describes the effect of diffusion mediated reactions 
between two atoms of the same specie
and the last two terms account for reactions between different species.
Each reaction rate is proportional to the number of pairs of atoms
of the two species involved, and to the sum of their sweeping rates.
The average population size    
of the $X_i$ specie on the grain is  
$\langle N_{i}\rangle = \sum_{N_1,N_2,N_3} N_i \ P(N_1,N_2,N_3)$,  
where 
$N_i=0,1,2 \dots$,
and
$i=1$, 2 or 3. 
The production rate 
per grain 
$R(X_k)$ 
(molecules s$^{-1}$)
of $X_k$ molecules
produced by the reaction
$X_i + X_j \rightarrow X_k$  
is given by 
$R(X_k) = (A_i+A_j) \langle N_i \  N_j \rangle$,
or by
$R(X_k)=A_i \langle N_i(N_i-1) \rangle$
in case that $i=j$.

In numerical simulations the master equation 
must be truncated in order to 
keep the number of equations finite.
A convenient way to achieve this is
to assign upper cutoffs
$N_i^{\max}$, $i=1,\dots,J$ 
on the population sizes,
where $J$ is the number of reactive species. 
The number of coupled equations
is thus
\begin{equation}
N_E = \prod_{i=1}^J (N_i^{\max}+1).
\label{eq:Nequations}
\end{equation}

\noindent
The truncated master equation is valid if the probability
to have larger populations beyond the cutoffs
is vanishingly small.
However, the number of equations, $N_E$, grows exponentially 
as the number of reactive species increases.
This severely limits the applicability of the master equation
to interstellar chemistry.
To exemplify the magnitude of this difficulty, 
note that the upper cutoffs must
satisfy
$N_i^{\rm max} \ge 1$
for the 
$J^{\prime}$ 
species that do not react with themselves
and
$N_i^{\rm max} \ge 2$ 
for the 
$J^{\prime \prime}$ 
species that do react with themselves.
Therefore, 
$N_E \ge 2^{J^{\prime}} 3^{J^{\prime \prime}}$
\cite{Stantcheva2002}.

The chemistry taking place on interstellar dust grains is
dominated by hydrogen atoms for three reasons. 
First, hydrogen is the most abundant specie;
Second, it hops on the surface much faster than other species; 
Third, it is the most
reactive specie, namely it reacts with many other species
which do not react with each other.
If two species such as $X_2$ and $X_3$ do not react with each other
it is not crucial to maintain the correlation between their
population sizes on a grain.
Therefore, the probability distribution of the population sizes
can be approximated by

\begin{equation}
P(N_1,N_2,N_3) = P(N_1) P(N_2/N_1) P(N_3/N_1),
\label{eq:independent}
\end{equation}

\noindent
where 
$P(N_i/N_1)$
is the conditional probability for a population size
$N_i$ of specie $X_i$, given that the population size
of $X_1$ atoms is $N_1$.

The multi-plane method is derived as follows:
inserting
Eq.
(\ref{eq:independent})
into the master equation
(\ref{eq:Master}),
we trace over $N_3$, using the fact that
$\sum_{N_3} P(N_3/N_1) = 1$
and
$\sum_{N_3} \dot P(N_3/N_1) = 0$.
We obtain the following set of equations:
\begin{eqnarray}  
&&\dot P(N_1,N_2) = 
\sum_{i=1}^2  
F_i \left[ P(..,N_i-1,..) - P(N_1,N_2) \right]  
\nonumber \\ 
&&+ \sum_{i=1}^2  
W_i \left[ (N_i+1) P(..,N_i+1,..) - N_i P(N_1,N_2) \right]  
\nonumber \\ 
&&+ \sum_{i=1}^2  
A_i \left[ (N_i+2)(N_i+1) P(..,N_i+2,..) 
           - N_i(N_i-1) P(N_1,N_2,N_2) \right]  
\nonumber\\ 
&&+ (A_1+A_2) \left[ (N_1+1)(N_2+1) P(N_1+1,N_2+1) 
           - N_1 N_2 P(N_1,N_2) \right] 
\nonumber\\ 
&&+ (A_1+A_3) \left[ (N_1+1) P(N_1+1,N_2) \langle N_3  \rangle_{N_1+1} 
           - N_1 P(N_1,N_2) \langle N_3  \rangle_{N_1}   \right], 
\label{eq:Master12}  
\end{eqnarray}  

\noindent
where
$\langle N_3\rangle_{N_1} = \sum_{N_3} N_3 P(N_3/N_1)$.
A similar procedure for tracing over $N_2$ yields
\begin{eqnarray}  
&&\dot P(N_1,N_3) = 
\sum_{i=1,3}
F_i \left[ P(..,N_i-1,..) - P(N_1,N_3) \right]  
\nonumber \\ 
&&+ \sum_{i=1,3}
W_i \left[ (N_i+1) P(..,N_i+1,..) - N_i P(N_1,N_3) \right]  
\nonumber \\ 
&&+ A_1 \left[ (N_1+2)(N_1+1) P(N_1+2,N_3) 
           - N_1(N_1-1) P(N_1,N_3) \right]  
\nonumber\\ 
&&+ (A_1+A_2) \left[ (N_1+1) P(N_1+1,N_3-1) \langle N_2  \rangle_{N_1+1} 
           - N_1 P(N_1,N_3) \langle N_2  \rangle_{N_1}  \right] 
\nonumber\\ 
&&+ (A_1+A_3) \left[ (N_1+1)(N_3+1) P(N_1+1,N_3+1) 
           - N_1 N_3 P(N_1,N_3) \right]. 
\label{eq:Master13}  
\end{eqnarray}  

\noindent
The production rates of the non-reactive species 
$X_4$, $X_5$ and $X_6$
are given by
$R(X_{i+3}) = A_i \sum_{N_1,N_2} N_i (N_i -1) P(N_1,N_2)$,
where $i=1,2$,
and
$R(X_6) = (A_1+A_3) \sum_{N_1,N_3} N_1 N_3 P(N_1,N_3)$.
The desorption rate of the reactive specie $X_3$ is given by
$R(X_3) = W_3 \sum_{N_1,N_3} N_3 P(N_1,N_3)$.
In simulations using the multi-plane method, one can choose
any initial condition that can be expressed by 
Eq. (\ref{eq:independent}).
A convenient choice is of an empty grain, 
namely,
$P(N_1=0,N_2=0)=1$
in Eq.
(\ref{eq:Master12})
and
$P(N_1=0,N_3=0)=1$
in Eq.
(\ref{eq:Master13}),
where all other probabilities vanish.
The multi-plane method requires
setting cutoffs,
$N_i^{\rm max}$, $i=1,2,3$,
where the same value of
$N_1^{\rm max}$
is used in Eqs.
(\ref{eq:Master12})
and
(\ref{eq:Master13}).

The simulations presented here were done for spherical grains
on which the density of adsorption sites is
$s = 5 \times 10^{13}$ 
(sites cm$^{-2}$), at a grain temperature
of $T=10$ K.
The activation energies for diffusion and desorption
of the reactive species 
H, O and OH were taken as
$E_0(1) = 22 $,
$E_1(1) = 32 $,
$E_0(2) = 25 $,
$E_1(2) = 32 $
and
$E_0(3) = 28 $,
$E_1(3) = 35 $ meV,
respectively.
The parameters for hydrogen are rounded values in the range
of experimental results on silicate and ice surfaces
\cite{Katz1999}.
For the other species no concrete experimental results are
available and the chosen values reflect the tendency of
heavier species to be more strongly bound to the surface.
The flux of H atoms on a grain of $S$ sites was 
taken as $F_1 = 2.75 \times 10^{-9} S$ (s$^{-1}$),
and the flux of O atoms was
$F_2 = 0.01 F_1$.
These parameters are suitable for dense molecular clouds
where such reaction networks are likely to take place.
For simplicity, the OH flux was neglected, taking $F_3=0$.
In Fig. 1 we present 
the production rates of H$_2$ (circles), O$_2$ (squares), H$_2$O 
(triangles) and
the desorption rate of OH ($\times$) per grain 
vs. grain size.
The results of the multi-plane method (symbols) 
and the complete 
master equation (solid lines),
both obtained by direct numerical integration,
are in excellent agreement.
The results of the rate equations, that for small grains deviate 
significantly from the master equation results
are also shown (dashed line).

In a complex reaction network of $J$
reactive species, 
$X_i$, $i=1,\dots,J$,
the distribution of surface populations is given by
$P(N_1,\dots,N_J)$.
If the only reactions are of
the form
$X_1 + X_i \rightarrow X_p$
and
$X_i + X_i \rightarrow X_q$
(no matter if $X_p$ and $X_q$ are reactive or not),
then the system can be reduced to 
$J-1$ sets of equations 
for
$P(N_1,N_j)$, $j=2,\dots,J$.
Each of these sets is obtained by tracing over
all the populations $N_i$ except for $N_1$ and $N_j$.
The number of equations
is reduced from
the exponential form 
of Eq.
(\ref{eq:Nequations})
to 
\begin{equation}
N_E = (N_1^{\rm max}+1) \sum_{j=2}^J (N_j^{\rm max}+1),
\end{equation}

\noindent
namely to a {\it linear dependence} on the number of reactive species.
In case that there is also a reaction between two different
species
$X_j + X_k \rightarrow X_p$,
where $j,k \ne 1$,
the three dimensional 
set of equations for
$P(N_1,N_j,N_k)$ 
should be  included in order to keep track of the 
correlations between these species.
The reaction network can be described by a graph in which
the nodes represent the reactive species, and any two
species that react are connected by an edge.
In general, each set of multi-plane equations is associated
with a maximal fully-connected subgraph, namely with a maximal set
of nodes which are all connected to each other.
Chemical networks tend to be sparse, namely most pairs
of species do not react. Therefore, the multi-plane equations
mostly consist of sets that invlove pairs of species and only 
few sets with three species or more.

As an example, consider the case in which CO molecules
are added to the H, O, OH system considered above
\cite{Stantcheva2002}.
This gives rise to the following sequence of hydrogen addition
reactions:
H + CO $\rightarrow$ HCO ($X_1 + X_7 \rightarrow X_8$),
H + HCO $\rightarrow$ H$_2$CO ($X_1 + X_8 \rightarrow X_9$),
H + H$_2$CO $\rightarrow$ H$_3$CO ($X_1 + X_9 \rightarrow X_{10}$)
and
H + H$_3$CO $\rightarrow$ CH$_3$OH ($X_1 + X_{10} \rightarrow X_{11}$).
Two other reactions that involve oxygen atoms also take place:
O + CO $\rightarrow$ CO$_2$ ($X_2 + X_7 \rightarrow X_{12}$)
and 
O + HCO $\rightarrow$ CO$_2$ + H ($X_2 + X_8 \rightarrow X_{12} + X_1$).
This reaction network is described by the graph shown in Fig. 2.
To account for the correlations between the populations of all pairs
of species that react with each other, the
multi-plane approach yields five sets of equations that
account for
$P(N_1,N_3)$,
$P(N_1,N_9)$,
$P(N_1,N_{10})$,
$P(N_1,N_2,N_7)$
and
$P(N_1,N_2,N_8)$.
The first three sets reside on planes while the last two 
sets are three dimensional sets that
are needed in order to account
for the correlations between the species that form CO$_2$.
Since the production of CO$_2$ is low compared to other products,
one may consider an approximation in which these correlations are
neglected, reducing the equations into only planar sets. 
In this case the set of equations will include:
$P(N_1,N_2)$,
$P(N_1,N_3)$,
$P(N_1,N_7)$,
$P(N_1,N_8)$,
$P(N_1,N_9)$
and
$P(N_1,N_{10})$.

In the simulations presented here
the activation energies for diffusion and desorption
of CO, HCO, H$_2$CO and H$_3$CO
were taken as
$E_0(7) = 30 $,
$E_1(7) = 36 $,
$E_0(8) = 30 $,
$E_1(8) = 38 $,
$E_0(9) = 33 $,
$E_1(9) = 39 $
and
$E_0(10) = 35 $,
$E_1(10) = 41 $ meV,
respectively.
The flux of CO molecules was taken as
$F_7 = 0.001 F_1$.
In Fig. 3 we present 
the production rates per grain of 
H$_2$O,
CO$_2$
and
CH$_3$OH
vs. grain size.
An excellent agreement is found between the multi-plane approach
(circles) and the complete master equation set (solid line).
For small grains the approximated multi-plane set ($+$) shows 
significant deviations in the production rate of CO$_2$ but
still provides good results for the other species.
The rate equation results are also shown (dashed line).
For this reaction network the number of equations required
in the complete master equation is between tens of
thousands to several millions, 
depending on the grain size. 
Numerical integration of the complete master equation is impractical. 
The steady-state
results presented here for the complete master equation set were obtained 
using Newton's method.
Unlike the complete set, the multi-plane set 
consists of up to about a thousand 
equations, and can be simulated efficiently 
by numerical integration 
both under steady state and time dependent conditions
using standard steppers such as Runge-Kutta. 

In summary, we have introduced the multi-plane method for the
simulation of the master equation of grain-surface chemistry. 
For complex reaction networks, this method
achieves a great reduction in 
the number of equations. 
It thus enables the incorporation of the master equation
in models of interstellar chemistry,
providing 
accurate results for the production rates of molecules on 
interstellar dust grains.
We expect that the multi-plane method will also be useful
in other problems that exhibit a related mathematical 
structure, such as the modeling of stratospheric cloud chemistry 
and the analysis of genetic networks in cells
\cite{McAdams1997,Paulsson2000}.

This work was supported by the Adler Foundation for Space Research
of the Israel Science Foundation.

%\bibliography{astrochemistry} 
%\bibliographystyle{prsty} 
 
\newpage 
\clearpage 
  
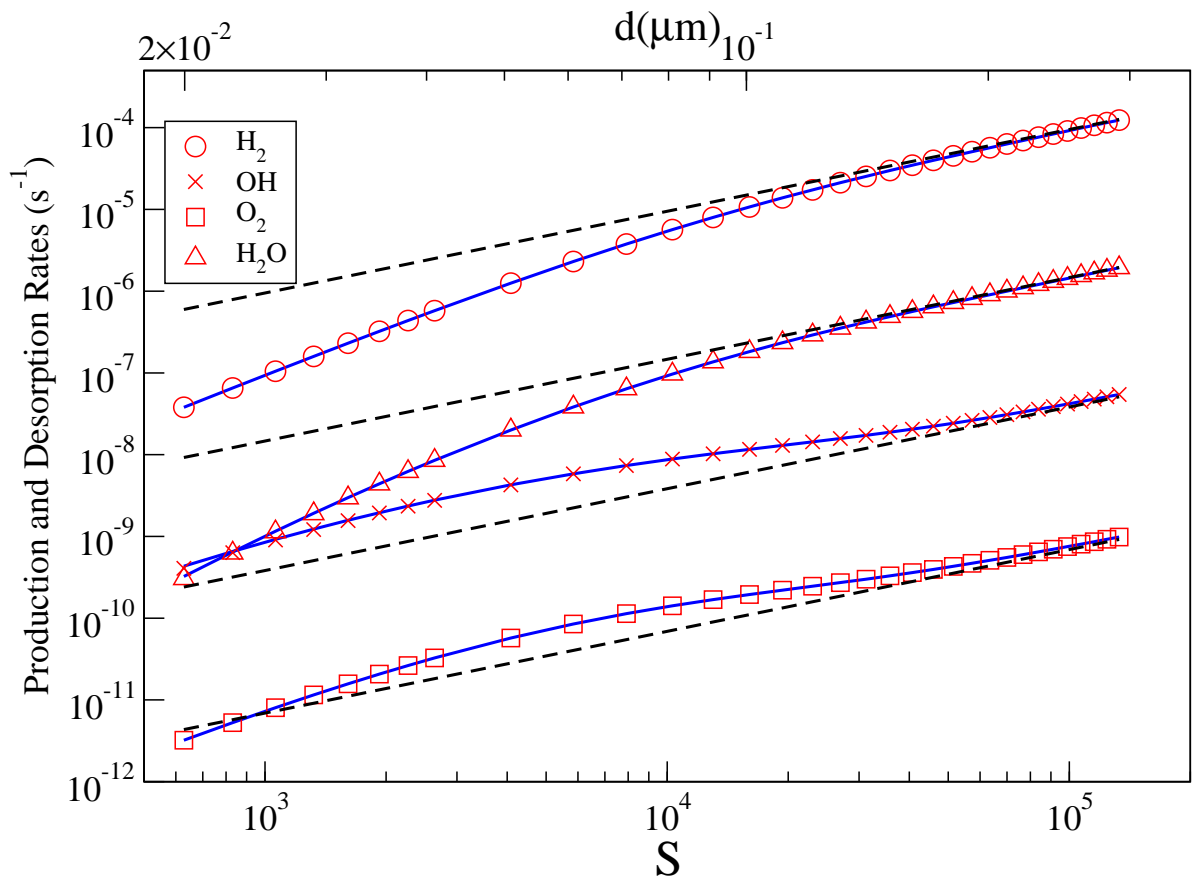
\begin{figure}  
\caption{
The production rates of H$_2$ (circles), O$_2$ (squares), H$_2$O 
(triangles) and
the desorption rate of OH ($\times$) per grain 
vs. grain size, given by the number of adsorption sites $S$
and the diameter $d$. Excellent agreement is found
between the multi-plane method (symbols) 
and the complete master equation (solid lines).
The results of the rate equations
are also shown (dashed lines).
}  
\label{fig:1}  
\end{figure}

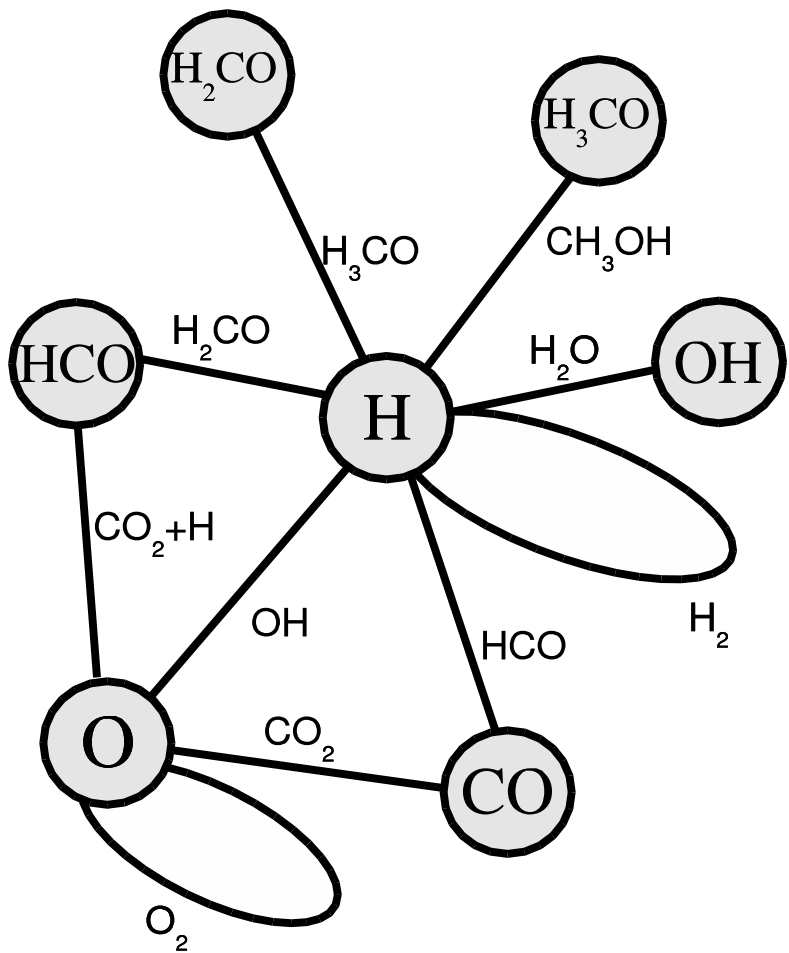
\begin{figure}  
\caption{
A graph describing the reaction network that results from
the adsorption of H and O atoms and CO molecules. The nodes 
represent the reactive species, while the edges connect pairs of
species that react with each other. The reaction products
are specified near the edges.
}  
\label{fig:2}  
\end{figure}

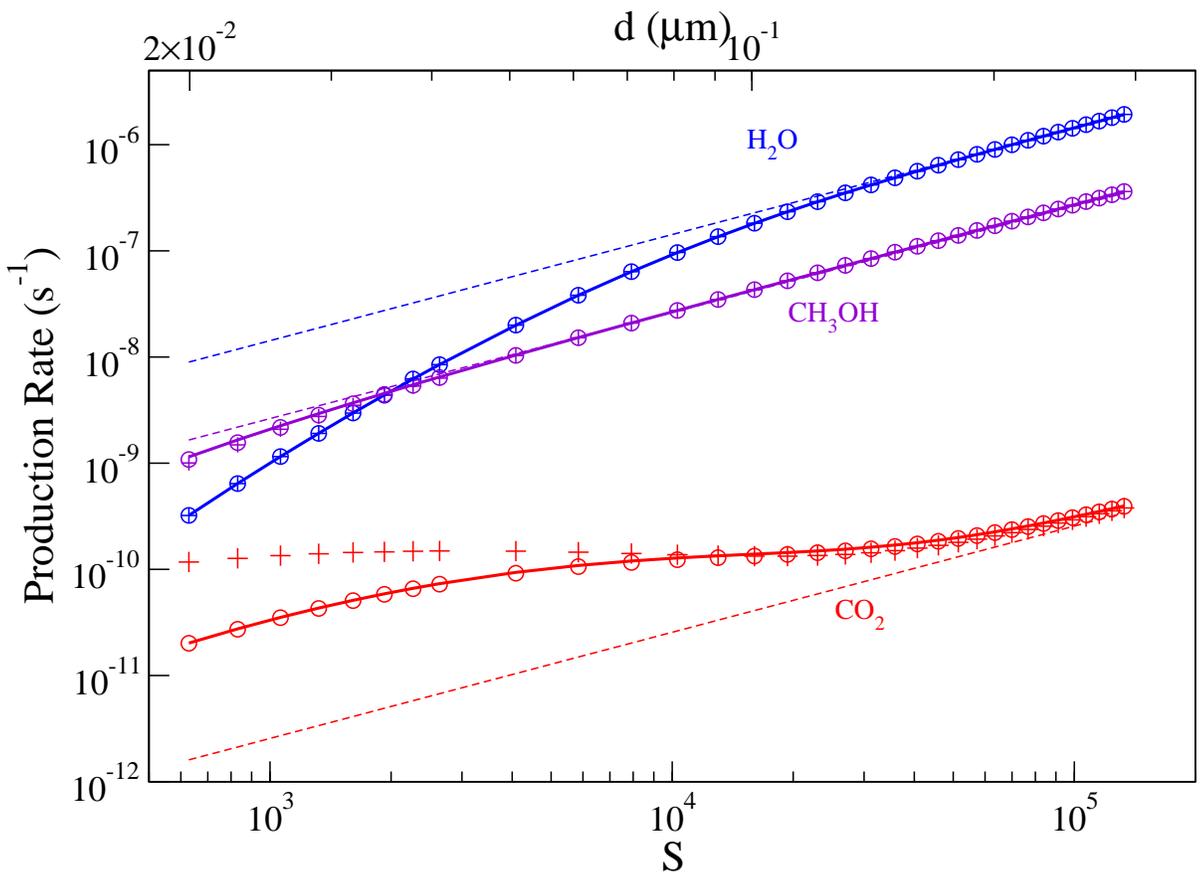
\begin{figure}  
\caption{
The production rates of 
H$_2$O,
CO$_2$
and
CH$_3$OH
vs. grain size as obtained from
the multi-plane approach
(circles) and the complete master equation (solid lines)
are found to be in excellent agreement.
The approximate multi-plane equations ($+$) 
deviate in the production rate of CO$_2$, but are still
accurate for the other species.
The rate equation results are also shown (dashed lines).
}  
\label{fig:3}  
\end{figure}  
  
\end{document}